\documentclass[twoside,11pt]{article}
\pdfoutput=1

\usepackage{amsthm,amsfonts,amsmath,amssymb,graphicx,url}
\usepackage{color}
\usepackage[ruled,vlined]{algorithm2e}
\usepackage{relsize}
\usepackage[margin=1in]{geometry}
\usepackage{fullpage}

\usepackage[T1]{fontenc}

\usepackage{hyperref}
\usepackage{cleveref}
\usepackage{enumitem}
\usepackage{subfig}

\usepackage{booktabs}
\usepackage{tikz}
\usepackage{natbib}
\usepackage{float}

\newcount\Comments  
\newcommand{\kibitz}[2]{\ifnum\Comments=1{\color{#1}{#2}}\fi}

\newtheorem{proposition}{Proposition}[section]
\newtheorem{theorem}[proposition]{Theorem}

\newtheorem{lemma}[proposition]{Lemma}
\newtheorem{claim}{Claim}

\theoremstyle{definition}
\newtheorem{definition}[proposition]{Definition}
\newtheorem{example}[proposition]{Example}
\newenvironment{Proof}
  {\begin{proof}}
  {\end{proof}}

\newcommand{\cM}{\mathcal{M}}
\newcommand{\cN}{\mathcal{N}}
\newcommand{\cB}{\mathcal{B}}

\newcommand{\cD}{\mathcal{D}}
\newcommand{\cG}{\mathcal{G}}
\newcommand{\cI}{\mathcal{I}}

\newcommand{\ee}{\mathbf{e}}
\newcommand{\vv}{\mathbf{v}}
\newcommand{\cc}{\mathbf{c}}

\newcommand{\odd}{{odd}}
\newcommand{\eve}{{even}}

\begin{document}

\title{
Your College Dorm and Dormmates:\\
Fair Resource Sharing with Externalities\thanks{The authors thank Edith Elkind for helpful discussions with her, and thank the anonymous reviewers for their insightful and constructive comments.
Part of this work was done when Bo Li was affiliated with the University of Oxford, and Yingkai Li was affiliated with Northwestern University. This work is funded by NSFC (No. 62102333), HKSAR RGC (No. PolyU 25211321) and CCF-Huawei Populus Grove Fund.
A preliminary version of this paper appeared in the Proceedings of the 19th International Conference on Autonomous Agents and MultiAgent Systems \cite{conf/atal/LiL20}. This full version, accepted in the Journal of Artificial Intelligence Research \cite*{dorm23}, includes all proofs that were omitted from the conference version as well as refined analysis and additional complexity results.  }}

\author{Jiarui Gan\thanks{Department of Computer Science, University of Oxford, United Kingdom. Email: \texttt{jiarui.gan@cs.ox.ac.uk}}
\and
Bo Li\thanks{Department of Computing, The Hong Kong Polytechnic University,
China. Email: \texttt{comp-bo.li@polyu.edu.hk}}
\and
Yingkai Li\thanks{Cowles Foundation for Research in Economics, Yale University, United States.
Email: \texttt{yingkai.li@yale.edu}}
}

\maketitle

\begin{abstract}
We study a fair resource sharing problem, where a set of resources are to be shared among a group of agents. Each agent demands one resource and each resource can serve a limited number of agents. An agent cares about what resource they get as well as the {\em externalities} imposed by their mates, who share the same resource with them. Clearly, the strong notion of envy-freeness, where no agent envies another for their resource {\em or} mates, cannot always be achieved and we show that even deciding the existence of such a strongly envy-free assignment is an intractable problem. Hence, a more interesting question is whether (and in what situations) a relaxed notion of envy-freeness, the {\em Pareto envy-freeness}, can be achieved. Under this relaxed notion, an agent envies another only when they envy both the resource {\em and} the mates of the other agent. 
In particular, we are interested in a dorm assignment problem, where students are to be assigned to dorms with the same capacity and they have dichotomous preference over their dormmates. We show that when the capacity of each dorm is 2, a Pareto envy-free assignment always exists and we present a polynomial-time algorithm to compute such an assignment. Nevertheless, the result breaks immediately when the capacity increases to 3, in which case even Pareto envy-freeness cannot be guaranteed.
In addition to the existential results, we also investigate the utility guarantees of (Pareto) envy-free assignments in our model.
\end{abstract}

\section{Introduction}

It is the back-to-school season, freshmen are starting their college life that they have long expected. Among the long list of things they are waiting to discover, they are all concerned about where they are going to live and who they are going to stay with in the forthcoming years.
This gives the accommodation administrator a hard time: whenever she proposes a dorm assignment, some student will come to complain that they prefer another dorm (due to the layout, location, surrounding environment, etc.), or prefer to stay with some other dormmates (due to the subjects they study, their hobbies, lifestyles, political views, etc.).
In this paper, we formalize the problem faced by the accommodation administrator as a fair resource sharing problem.
In this problem, a set of heterogeneous resources needs to be fairly shared among a group of agents who demand one resource each. The agents have preferences over both the resources and their mates, i.e., the other agents with whom they share the same resources. Each resource can be assigned to multiple agents, subject to its capacity, i.e., the maximum number of agents it can accommodate.
The aim is to find an {\em envy-free (EF)} \cite{foley1967resource} assignment of the resources to the agents, such that no agent would prefer to exchange their resource with any other agent. 
Besides dorm assignment, fair resource sharing problems also arise in many other scenarios, such as project assignment, where each agent has their own preference over projects as well as collaborators; or group activity selection problem, where agents have preferences over activities as well as group mates.

Accordingly, in this work, we propose a resource sharing model where the agents care about what resource they get as well as the {\em externalities} imposed by their mates, who share the same resource with them.
We investigate the fair resource allocation problem in this model from the perspective of algorithm design and computational complexity. 

The concept of externalities in the context of fair allocation is not new and dates back to as early as the work of \citet{moulin1990uniform}, or even earlier study on stable matching \cite{gale1962college,irving1985efficient}, that are in a similar vein.
Nevertheless, subtle differences between the existing models and ours make the results and techniques required to derive the results drastically different.
For example, externalities have been considered in the private resource allocation model where the resources cannot be shared among the agents \cite{mishra2021fair}.
The aforementioned stable matching models and hedonic games can also be viewed as models with sharable resources \cite{bogomolnaia2002stability,elkind2009hedonic}, but in this sense the agents only care about who they share resources with but not about the values of the resources (e.g., the resources are homogeneous).
To the best of our knowledge, no results for these models would imply ours in this paper. We will compare our work and related work in more detail in Section~\ref{sec:related-works}.

\subsection{Main Results}

We first study a strong EF notion where every agent considers the total value of their resource and mates and does not envy another.
We show that this strong EF notion does not ensure the existence of an EF assignment, and deciding the existence of an EF assignment is NP-hard. The hardness result holds even when there are only two resources and all the agents have dichotomous preferences over the other agents.

Indeed, EF assignments seldom exist under the above strong notion even in very simple settings, so we are interested in a relaxed notion called {\em Pareto envy-freeness (PEF)}.
Informally speaking, an assignment is called PEF if for any two agents $i$ and~$j$,
either $i$ does not envy $j$ for the resource $j$ receives,
or for the {\em mates} of $j$ that share the same resource with $j$.
With this relaxed notion, we show additional negative and positive results.

Specifically, a PEF assignment, still, may not exist and it is NP-hard to decide the existence; nevertheless, when we focus on a special class of the resource sharing problem where the resources have the same capacity and agents have dichotomous preference over their mates, we find that a PEF assignment always exists if the resource capacity is 2 and  such an assignment can be found in polynomial time. 
The proof of this result is non-trivial and relies on Gallai-Edmonds Theorem \cite{lovasz2009matching} and Hall's Theorem \cite{hall1934representation}.
The algorithm computing a PEF allocation when the capacity is 2 is regarded as the main technical contribution of this work.
We also show that the capacity of $2$ is a tight upper bound of the resource capacity that guarantees the existence of a PEF assignment; when the capacities increase to 3, we present a counterexample that admits no PEF assignment.

We also investigate the utility implications of EF and PEF allocations. 
We first find that an EF allocation may not able to ensure every agent's utility to be no smaller than the average of her total utility. 
Nevertheless, an approximate version holds.
We show that every EF assignment ensures at least a half of the average value as long as the capacity of each resource is at least 2, and the bound of the approximation ratio is tight.
In addition, PEF also implies similar properties under a Pareto optimality-style notion.

\subsection{Organization of the Paper}

The remainder of this paper is organized as follows.
In Section~\ref{sec:related-works}, we first provide a detailed review of related work in the literature.
In Section~\ref{sc:model}, we describe the resource sharing model and a special setting of the model called the {\em dorm sharing model}. Then, we formalize our EF and PEF notions in Section~\ref{sc:ef-notions} and discuss the existence of an assignment under these notions, as well as the related computational complexity results. 
In Section~\ref{sec:dorm}, we study PEF assignments in dorm sharing models and present a key result --- a polynomial-time algorithm to compute a PEF assignment when the dorm capacity is 2. 
In Section~\ref{sc:EF-PROP}, we further investigate the utility guarantees of EF and PEF assignments. 
We conclude in Section~\ref{sc:conclusion} where we discuss directions for future work and alternative ways to define the EF notions.

\section{Related Work}
\label{sec:related-works}


Our work is related to several lines of research in the literature but with the following key differences.
First, our resource sharing model allows one resource to be allocated to multiple agents, thus differing from models studied in the private resource allocation literature.
Second, in our model, every agent cares about both the allocated resource and other agents who share the same resource with them; hence, it is different from cooperative game theoretic models, such as hedonic games, and stable matching. 
Finally, unlike fair rent division problems, monetary transfers among the agents are not allowed in our model.
We review the related work in more detail below.

\paragraph{Envy-free Resource Allocation}
The primary consideration of our work, that of EF, is related to the substantial literature on fair division problems when the resources cannot be shared among the agents.
The major settings of fair division problems include:
(i) the division of a {\em divisible} good, e.g., a cake, that can be thought of as a segment that can be cut through at any points, hence also known as the {\em cake-cutting} problem \cite{steihaus1948problem,dubins1961cut,brams1996fair,aziz2016discrete};
(ii) the allocation of a set of {\em indivisible} goods, each can only be assigned to some agent as a whole \cite{lipton2004approximately,budish2011combinatorial,caragiannis2019unreasonable}; and
(iii) the inverse problems of allocating divisible or indivisible {\em bads}, i.e., items that come with a cost for the agents, such as chores  \cite{aziz2017algorithms,aziz2019weighted,huang2019algorithmic}.
Envy-freeness and proportionality are the main considerations in the studies of these problems.
Although an envy-free allocation and a proportional allocation normally always exist in the cake-cutting setting, they may not exist when the items are indivisible. Thus, relaxed notions are proposed. Typically, {\em envy-freeness up to one item}, or {\em EF-1}, is proposed as a relaxation of envy-freeness \cite{lipton2004approximately} and \cite{budish2011combinatorial}, which can be guaranteed with indivisible items and is hence a pervasive notion in studies of allocations of indivisible items. Similarly, the {\em maximin share fairness (MMS-fairness)} is proposed \cite{budish2011combinatorial}, which requires that each agent obtains at least the maximum value they can get by partitioning the items into $n$ bundles and taking the least valuable bundle; it has been shown that constant approximations of the MMS-fairness can be achieved with indivisible items \cite{DBLP:journals/talg/AmanatidisMNS17,DBLP:journals/jacm/KurokawaPW18,DBLP:conf/sigecom/GargT20}.
Externalities are considered in both divisible and indivisible resource allocation models. 
For divisible resources, \citet{DBLP:conf/ijcai/BranzeiPZ13} extended the fairness notions to the setting where the agents' utilities are not only decided by their own pieces, but also by the pieces of other agents.
\citet{DBLP:conf/www/SeddighinSG19} and \citet{mishra2021fair} further studied similar problems when resources are indivisible.
\citet{DBLP:conf/ijcai/ElkindPTZ20} studied a mixed utility function in land allocation scenarios which combines the utility for the land and the externality for nearby friends.
In all the above works, the resources are restricted to be not sharable.
Instead, a parallel line of research investigates the problem when the resources can be shared among groups of agents, where the group formation can be prefixed or arbitrary, and group fairness (e.g.,  weighted envy-freeness) is studied accordingly;
see, e.g., \cite{DBLP:journals/mss/ManurangsiS17,DBLP:journals/scw/Segal-HaleviN19,DBLP:journals/tcs/KyropoulouSV20,DBLP:journals/teco/ChakrabortyISZ21}.

\paragraph{Cooperative Games}
In most cooperative game theoretic models, such as hedonic games \cite{bogomolnaia2002stability,elkind2009hedonic,aziz2019fractional}, 
only the external values for the mates are considered. 
\citet{skibski2020fair} studied the fair allocation of payoffs in cooperative games with externalities, where agents form coalitions and the value of each coalition is defined not only by the members of the coalition but also by the other coalitions; however, there is no resource shared among the agents. 
One exception is the {group activity selection game} \cite{darmann2012group,darmann2015group,eiben2018structural}, where agents are assigned to different group activities and their utilities depend on the type of activity (similar to resources in our model) as well as the members in the same group. 
Despite the similarity, group activity selection games usually employ different modeling assumptions,
e.g., every agent always has the choice of deviating to singleton activities (i.e., by being alone), which is not feasible in our model, or the supply of activities are not fixed, unlike that of the dorms.
Meanwhile, the main objective of group activity selection games is usually to incentivize agents to form stable groups or to participate in (non-singleton) activities whereas our focus is on EF.

\paragraph{Stable Matching}
Initiated by \citet{gale1962college}, the {\em stable roommate problem} has been studied extensively. In this problem, $2m$ students are to be assigned to $m$ dorms in a way such that no pair of students want to swap their positions in the assignment, so that the assignment is considered to be {\em stable}. 
There are also other stability notions, such as exchange stability \cite{cechlarova2005exchange,DBLP:conf/sagt/Bodine-BaronLCHW11} and popular matching \cite{biro2010popular} among many others \cite{kojima2013matching,pycia2021matching}.
These papers only consider the agents' preferences over their mates. The dorms are assumed to be identical to every agent. Though we can generalize this stability notion by incorporating also the agents' values for the dorms, a stable dorm assignment can be far from being EF. Arguably, in some scenarios, such as the dorm assignment scenario, EF is a more suitable notion than swap stability as it offers better fairness.
A more specific setting, called the {\em fair house assignment problem} and initiated by \cite{hylland1979efficient,shapley1974cores} (and more recently, \cite{gan2019envy,beynier2019local,aigner2019envy,DBLP:conf/atal/Gross-HumbertBB21}), is a special case of our model where the capacities of the resources are one (and external values between agents are zero).
We adopt EF as our key consideration and adapt the notion to our setting by considering the Pareto EF.
A similar relaxed notion of EF was also considered by \citet{chan2016assignment} and in a follow-up work by \citet{huzhang2017online}. They showed that when the capacity of the dorms increases to 2, although individual EF cannot be guaranteed, a direct application of the result in \cite{shapley1971assignment} ensures {\em room EF}, a notion that treats every pair of dormmates as a whole. Namely, by moving all the agents in a dorm to another dorm, the agents' total utility cannot be increased. This is different from our objective, where we are concerned with the utility of each individual agent, rather than that of a group.

\paragraph{Fair Sharing with Money}
Similar fair sharing problems have also been studied in various market settings. 
In a market, resources have (different) prices (e.g., rent) that need to be shared between the agents that they are assigned to. Hence, the utility of each agent is the net value they obtain, i.e., the difference between the value they obtain from the resource and the price they pay.
As shown by Shapley and Shubik \cite{shapley1971assignment}, if each resource has capacity 1,
then there is an assignment along with a price profile, such that the matching between the agents and resource is envy-free.
In our model, however, we consider only scenarios where monetary transfer is {\em not} possible.
The counterpart of this setting is related to the {\em fair rent division} and {\em rental harmony} problems \cite{alkan1991fair,aragones1995derivation,edward1999rental,abdulkadirouglu2004room,gal2016fairest,DBLP:conf/cocoa/GhodsiLMMS18}, which studies fair ways to assign rooms to agents and divide the rent among them. With monetary transfers, an envy-free solution is always feasible; thus one research interest is to find the ``best'' envy-free solutions \cite{alkan1991fair,gal2016fairest}. Apart from allowing monetary transfer, these papers also differ from our work in that they did not consider the agents' preferences over their dormmates. Indeed, when these external preferences are considered, the setting with monetary transfer is an interesting parallel direction; for example, \citet{velez2016fairness} have considered linear externalities in their work.

\section{Resource Sharing Model}
\label{sc:model}

There is a set $M$ of resources (e.g., dorms) that need to be assigned to a set $N$ of agents.
Let $m$ and $n$ be the sizes of $M$ and $N$, respectively, i.e., $|N| = n$ and $|M| = m$.
Each agent $i\in N$ has a {\em value} $v_{ij} \ge 0$ for each resource $j \in M$ and demands exactly one resource.
Besides the values for the resources, each agent also gets a value from their {\em mates}, i.e., the set of agents who share the same resource with them.
To distinguish between these two types of values, we will refer to an agent's value for a resource as an {\em internal value} and their value for another agent as an {\em external value} or an {\em externality}.
We consider additive values throughout this paper: agent $i$ receives an external value $e_{i j} \in \mathbb{R}$ if she shares her resource with agent~$j \in N$; the total external value she receives is defined to be $e_i(A) = \sum_{j \in A} e_{i j}$ if she shares a resource with agents $A \subseteq N$.
We assume that $e_{ij} \ge 0$ for all $i,j \in N$, and $e_{ii} = 0$ (i.e., agents do not have external value for themselves).

In a feasible assignment, each agent is assigned to one resource, while each resource can serve multiple agents. We further set a capacity $c_j \ge 1$ for each resource $j \in M$, which is the maximum number of agents $j$ can serve. Without loss of generality, we assume that the total supply meets the total demand, so we have $n = \sum_{j\in M} c_j$.
An instance of the resource sharing problem is then given by a tuple $\cI = (N,M,\vv,\ee,\cc)$, where  $\cc = (c_j)_{j\in M}$, $\vv = (v_{ij})_{i\in N, j\in M}$, and $\ee = (\ee_{ij})_{i,j\in N}$.

We write an assignment as $X = (X_1, \dots, X_m)$, where each $X_j \subseteq N$ denotes the subset of agents assigned to resource $j$. For each agent $i\in N$, we denote by $r_i(X)$ the resource assigned to $i$ in $X$, and by $S_i(X)$ the set of agents sharing the same resource with $i$ in $X$.
With slight abuse of notation, the internal and external values each agent $i$ obtains from assignment $X$ are denoted as $v_i(X)$ and $e_i(X)$, respectively;
we have 
$$v_i(X) = v_{i r_i(X)}$$
and
$$e_i(X) = \sum_{\ell \in S_i(X)} e_{i \ell}.$$

\paragraph{Dorm Sharing Model}

We are particularly interested in a special setting of the above model, which we call the {\em dorm sharing model} as it offers a basic model for dorm assignment tasks.
In a dorm sharing model, the following conditions hold:
\begin{enumerate}
\item All resources (i.e., dorms) have the same capacity $c \ge 1$ (so $n = c\cdot m$);
\item Every agent has a dichotomous preference over the other agents, i.e., their external values are binary: $e_{ij} \in \{0,1\}$ for all $i,j \in N$;
\item The external values are symmetric, i.e. $e_{ij} = e_{ji}$ for all $i,j \in N$.
When $e_{ij} = e_{ji} = 1$, we say that agents $i$ and $j$ are {\em friends} of each other.  
\end{enumerate}

Thus in the dorm sharing model, the externalities among the agents can be described as an undirected graph $G=(N,E)$, where each node represents an agent and there is an edge $e= \{i,j\} \in E$ between two agents $i$ and $j$ if and only if they are {\em friends} of each other.
We refer to such a graph as an {\em externality graph}.
The corresponding fair dorm assignment problem asks how to find a fair (EF, in particular) assignment in the dorm sharing model. 
We define the fair notions next and analyze the existence of fair assignments under these notions.

\section{EF Notions and Existence of EF Assignments}
\label{sc:ef-notions}

Our goal is to find an EF assignment, in which no agent would envy another for what they are assigned, considering both the internal and external values.
Arguably, the most straightforward approach is to consider the sum of the internal and external values. 
We define the (total) utility of an agent $i$ in an assignment $X$ as 
$$u_i(X) = v_i(X) + e_i(X).$$
Note that 
the above definition of utilities is as general as one that allows personalized weights for internal and external values.
That is, if we have $u_i(X) = w_i\cdot v_i(X) + e_i(X)$ for some $w_i \ge 0$, we can consider an equivalent instance of our model where $v'_{ij} = w_i \cdot v_{ij}$ for all $j \in M$.
The envy-free notion defined below expects that no agent envies another agent for what they get at their position.

\newcommand{\exch}{\leftrightarrow}

\begin{definition}[\bf EF assignment]\label{def:ef}
For any assignment $X$, let $X^{i \exch j}$ be the assignment resulting from switching agents $i$ and $j$ in $X$, i.e., $r_i(X) = r_j(X^{i \exch j})$, $r_j(X) = r_i(X^{i \exch j})$, and $r_\ell(X) = r_\ell(X^{i \exch j})$ for any $\ell \notin \{i, j\}$.
An assignment $X$ is {\em envy-free (EF)} if $u_i(X) \ge u_i(X^{i \exch j})$ for every pair of agents $i,j \in N$.
\end{definition}

Unfortunately, an EF assignment may not exist in a resource sharing instance, even in the special case of dorm sharing.  For example, when all agents have zero external values and they all prefer dorm $1$ to any other dorms, any agent who is not assigned dorm $1$ will envy those who get it.
Moreover, even deciding the existence of an EF assignment appears to be computationally hard as we show in Proposition~\ref{thm:EF-NP-c}.
Because of these negative results, we consider a relaxed EF notion, which we term the {\em Pareto EF (PEF)}. 
The PEF notion treats an agent's utility as a two-dimensional vector whose components are the internal and external utilities; 
a utility vector is considered to be {\em no} worse than another vector if it is not dominated by that vector in both dimensions (Definition~\ref{def:PEF}). 

\begin{proposition}
\label{thm:EF-NP-c}
Deciding whether a given dorm sharing instance admits an EF assignment is NP-complete, 
even when there are only two resources that are of value~$0$ for every agent.
\end{proposition}

\begin{Proof}
The problem is clearly in NP: an EF assignment serves as a witness of a yes-instance; the EF of this assignment can be verified in polynomial time.
To prove the NP-hardness, we present a reduction from the well-known NP-complete problem, the {\sc Clique} problem.
An instance of {\sc Clique} is given by an undirected graph $G = (V,E)$ and an integer $k > 0$. It is a yes-instance if and only if there exists a {\em clique} of size $k$ on $G$, i.e., a subset $X \subseteq V$, such that $\{i, j\} \in E$ for every pair of distinct vertices $i, j \in X$.
Without loss of generality, we can assume that $ |V|/2 < k \le |V|$:
when $k \le |V|/2$, we can always modify an instance to one that satisfies this assumption by adding $|V|$ dummy vertices to the graph, that form a clique and are connected to all vertices in $V$;
the new instance $\langle \tilde{G} = (\tilde{V}, \tilde{E}),\, \tilde{k} = k + |V| \rangle$ is such that $\tilde{k} \ge |\tilde{V}|/2$ and it is a yes-instance if and only if the original instance $\langle G, k \rangle$ is a yes-instance.

For ease of presentation, we first prove this result for a model where \emph{not} every agent has value $0$ for all the resources, and then show how it can be extended to the situation where every agent has value $0$ for all the resources.

Given an instance $\langle G, k \rangle$ of {\sc Clique}, we construct an instance of our problem as follows.
Let the set of agents be $N = N_1 \cup N_V \cup N'_V$, where:

\begin{itemize}
\item
$N_1$ contains a set of $4k - 2|V|$ agents who have value $1$ for resource $1$, and value $0$ for resource $2$ and every other agent. Thus, in an EF assignment, these agents have to be assigned to resource $1$.

\item
$N_V = \{a_i : i \in V\}$ contains $|V|$ agents, each corresponding to a vertex on $G$.
We say that two agents $a_i, a_j \in N_V$ are friends if and only if $\{i,j\} \in E$.
Each agent $a_i \in N_V$ has value $2k - d_i - 1$ for resource $1$, external value $1$ for each of their friends in $N_V$, and external value $1$ for $a'_i \in N'_V$ defined below, where $d_i$ denotes the degree of $i$ on $G$ (i.e., the number of friends in $N_V$). All their other values (including external values) are $0$.

\item
$N'_V = \{a'_i : i \in V\}$ contains $|V|$ agents, each corresponding to a vertex on $G$. Each $a'_i \in N'_V$ has external value $1$ for their counterpart $a_i \in N_V$. All their other values (including external values) are $0$.
Thus, in an EF assignment, each $a'_i$ has to be assigned to the same resource $a_i$ is assigned to; we should always allocate them as a pair.
\end{itemize}
There are $4k$ agents in total, so we set $c_1 = c_2 = 2k$.
We next argue that the {\sc Clique} instance is a yes-instance if and only if the above instance admits an EF assignment.

First, suppose that the {\sc Clique} instance is a yes-instance: there exists a clique $X\subseteq V$, $|X| = k$. Consider the following assignment.
Assign all agents in $N_1$ to resource $1$;
assign each $a_i \in N_V$, $i \in X$, to resource $2$ and the remaining agents in $N_V$ to resource $1$;
assign each $a'_i \in N'_V$ to the same resource $a_i$ is assigned to.
The assignment satisfies the capacity constraint and it is EF:
\begin{itemize}
\item
It is EF for every agent in $N_1$ and $N'_V$ by our observation above.

\item
Each $a_i \in N_V$, $i \in X$, gets external utility $k$ from $a'_i$ and their $k-1$ friends on resource $2$; they have $d_i - k + 1$ remaining friends on resource $1$, so swapping them to resource $1$ gives them utility at most $(2k - d_i - 1) + (d_i - k + 1) = k$. The assignment is EF for them.

\item
Each $a_i \in N_V$, $i \notin X$, gets external utility at least $d_i - k + 1$ from $a'_i$ and at least $d_i - k$ friends on resource $1$, so their utility is at least $(2k - d_i - 1) + (d_i - k + 1) = k$. Swapping them to resource $2$ gives them at most $k$ friends while they have value $0$ for the resource, so the assignment if EF for them.
\end{itemize}

Conversely, suppose that there exists no clique of size $k$ on $G$, and for the sake of contradiction, there exists an EF assignment.
Again, by our observation, all agents in $N_1$ have to be assigned to resource $1$, and each pair $a_i$ and $a'_i$ has to be assigned to the same resource. Hence, $k$ pairs of $a_i$ and $a'_i$ are assigned to resource $2$. When there exists no size-$k$ clique on $G$, some $a_i$ on resource $2$ finds at most $k-2$ friends on the same resource, obtaining utility at most $k-1$. There are at least $d_i - k + 2$ friends of this agent on resource $1$, so swapping them to resource $1$ (with some agent in $N_1$) gives them utility at least $(2k - d_i - 1) + (d_i - k + 2) > k - 1$, which contradicts the assumption that the assignment is EF.

\smallskip
We have finished the proof for the setting where agents may have non-zero resource values.
To extend the proof to the setting where every agent has value $0$ for every resource, the idea is to convert non-zero resource values in our reduction to external values by adding extra agents. We briefly describe the approach below.
\begin{itemize}
\item
We let $N_1$ now contain $4k + 2|V|$ agents who have external value $1$ for each other; let $N_2$ be a new set of $4|V|$ agents who have external value $1$ for each other.
The idea is to ensure that all agents in $N_1$ (or $N_2$) must be assigned to the same resource if we want the assignment to be EF.
Now that we have $8|V|$ more agents, we set $c_1 = c_2 = 2k + 4|V|$; this further ensures that $N_1$ and $N_2$ must be put on different resources as the capacity of one resource is not large enough to hold both groups simultaneously, so we end up having $2k$ free space on one resource and $2|V| - 2k$ on another, which is equivalent to our reduction above.
It then only matters how we distribute the remaining agents in these free spaces.

\item
We let each agent $a_i \in N_V$ have external value $1$ with each of the first $2k - d_i - 1$ agents in $N_1$, so given that all agents in $N_1$ should be on the same resource in an EF assignment, $a_i$ will gain utility $2k - d_i - 1$ by moving to this resource, which is equivalent to their value for resource $1$ in our reduction above. Note that this does not change the fact that all agents in $N_1$ must be on the same resource, even though now they can also gain external utility from agents in $N_V$: observe that if some agent in $N_1$ is assigned to the same resource $N_2$ is assigned to, then this agent gets external utility at most $2k-1$; in contrast, the agent is able to gain an external utility of at least $2k + 2|V|$ if they move to the other resource as there must be at least this many agents in $N_1$ assigned to the other resource.
\end{itemize}
This completes the proof.\footnote{The setting where agents have value $0$ for the resources share some similarities with the following {\em internal partition} problem: given a graph $G=(V,E)$ and a ratio $q \in [0,1]$, is there a partition of $V$ such that every vertex has at least as $q$ fraction of neighbours in its own part \cite{ban2016internal}. It is shown that internal partition is NP-complete for some specific ratios \cite{bouquet2020partition}. Our problem is similar to the case where $q = 1/2$ but with a subtle difference due to our consideration of swap stability (which means that the ratio is not necessarily at least $1/2$ for every agent in an EF assignment). Nevertheless, our reduction directly implies the NP-hardness of internal partition with $q = 1/2$ and the additional requirement that the two parts in the partition are of equal size.}
\end{Proof}

\begin{definition}[\bf PEF assignment]
\label{def:PEF}
An assignment $X$ is {\em Pareto envy-free (PEF)} if for every pair of agents $i$ and $j$ at least one of the following two conditions holds:
\begin{enumerate}
\item $v_i(X) \ge v_i(X^{i\exch j})$; or
\item $e_i(X) \ge e_i(X^{i \exch j})$.
\end{enumerate}
\end{definition}

By definition, EF implies PEF.
Intuitively, the PEF notion assumes that an agent is happy if she finds no other agent with a better resource and a better set of mates in the assignment.
While comparing the internal and external values separately weakens the EF notion, it might be a more appropriate approach for some scenarios where these two types of values are not directly comparable.
For example, the internal value may represent the negative of the monetary cost (e.g., rent) of a resource, while the external value may represent friendship, which cannot always be converted to monetary values.
Likewise, the PEF notion also applies when the agents only have cardinal utilities for the resources and ordinal preferences over sets of mates, or vice versa.

It turns out that the relaxation of EF to PEF does not immediately free us from ``existential crises'': a PEF assignment may not exist even in the dorm sharing model and even when there are only two dorms; we illustrate this via Example~\ref{example:no-PEF} and Proposition~\ref{prp:example-no-PEF}.
Similarly, deciding the existence of a PEF assignment is computationally hard in a slightly more general setting where resources are allowed to have different capacities; we present Proposition~\ref{thm:PEF-NPhard}.\footnote{We are only able to show the hardness result for the situation where resources have different capacities, and we leave the complexity when resources are restricted to have the same capacity as an interesting open problem.}
Despite the string of negative results and theoretical barriers, our key finding in this paper is that a PEF assignment always exists if all the dorms have a capacity of $2$, and such an assignment can be computed in polynomial time.
We show these results in the next section.

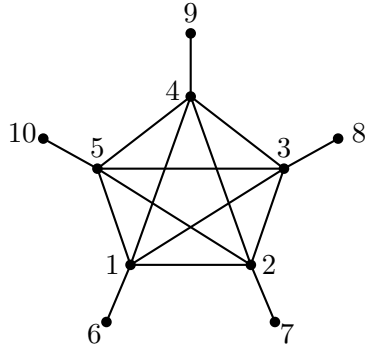
\begin{figure}[htbp]
\begin{center}
\vspace{3mm}
\setlength{\unitlength}{0.6cm}
\begin{tikzpicture}[scale = 0.8]

\begin{scope}[thick]

\filldraw[fill=black, draw=black] (3, 1) circle (0.07cm);
\filldraw[fill=black, draw=black] (5, 1) circle (0.07cm);
\filldraw[fill=black, draw=black] (2.45, 2.6) circle (0.07cm);
\filldraw[fill=black, draw=black] (5.55, 2.6) circle (0.07cm);
\filldraw[fill=black, draw=black] (4, 3.8) circle (0.07cm);
\draw (3, 1) -- (5, 1);
\draw (3, 1) -- (2.45, 2.6);
\draw (3, 1) -- (5.55, 2.6);
\draw (3, 1) -- (4, 3.8);

\draw (5, 1) -- (2.45, 2.6);
\draw (5, 1) -- (5.55, 2.6);
\draw (5, 1) -- (4, 3.8);

\draw (2.45, 2.6) -- (5.55, 2.6);
\draw (2.45, 2.6) -- (4, 3.8);

\draw (5.55, 2.6) -- (4, 3.8);

\filldraw[fill=black, draw=black] (4, 4.85) circle (0.07cm);
\draw (4, 4.85) -- (4, 3.8);
\filldraw[fill=black, draw=black] (1.55, 3.1) circle (0.07cm);
\draw (1.55, 3.1) -- (2.45, 2.6);
\filldraw[fill=black, draw=black] (6.45, 3.1) circle (0.07cm);
\draw (6.45, 3.1) -- (5.55, 2.6);
\filldraw[fill=black, draw=black] (2.6, 0.05) circle (0.07cm);
\draw (2.6, 0.05) -- (3, 1);
\filldraw[fill=black, draw=black] (5.4, 0.05) circle (0.07cm);
\draw (5.4, 0.05) -- (5, 1);

\draw (3.7, 3.9) node {$4$};
\draw (2.45, 2.95) node {$5$};
\draw (2.7, 1) node {$1$};
\draw (5.3, 1) node {$2$};
\draw (5.55, 2.95) node {$3$};
\draw (4, 5.2) node {$9$};
\draw (1.2, 3.2) node {$10$};
\draw (2.4, -0.15) node {$6$};
\draw (5.6, -0.15) node {$7$};
\draw (6.8, 3.2) node {$8$};

\end{scope}

\end{tikzpicture}
\vspace{3mm}
\caption{The externality graph of Example~\ref{example:no-PEF}.}
\label{figure:example:restrict}
\end{center}
\end{figure}

\begin{example}
\label{example:no-PEF}
There are 2 dorms both of capacity 5, and 10 agents $N= \{1,\dots, 10\}$:
\begin{itemize}
\item
 each agent in $\{1,\dots, 7\}$ has value $1$ for dorm 1 and value $0$ for dorm 2;
 \item
 each agent in $\{8,9,10\}$ has value $0$ for dorm 1 and value $1$ for dorm 2.
\end{itemize}
All agents in $\{1,\dots,5\}$ are friends with each other; and for each $i \in \{1,\dots,5\}$, agent $i$ and $i+5$ are friends (see the externality graph in Figure \ref{figure:example:restrict}).
\end{example}


\begin{proposition}
\label{prp:example-no-PEF}
The instance in Example~\ref{example:no-PEF} does not admit any PEF assignment.
\end{proposition}

\begin{Proof}
Let $N^* = \{1,\dots, 5\}$.
We consider all possible assignments of the agents in $N^*$.

\begin{itemize}
\item
Case 1. All agents in $N^*$ are assigned to dorm 1. 
In this case, all the other agents in $\{6,\dots,10\}$ are assigned to dorm 2. 
In this assignment, agent 6 does not share a dorm with her friends and is assigned to the worse dorm.
Thus, agent 6 Pareto-envies every agent in $N^* \setminus\{1\}$.

\item
Case 2. All agents in $N^*$ are assigned to dorm 2.
Similar to Case 1, agent 8 Pareto-envies every agent in $N^* \setminus\{3\}$ in this assignment.

\item
Case 3. Four agents in $N^*$ are assigned to dorm 1 and the other one is assigned to dorm 2. 
Suppose agent $i \in N^*$ is the one assigned to dorm 2.
Other than the four agents in $N^*$, there is another agent $j \notin N^*$ who is assigned to dorm 1. 
For agent $i$, at least four of her friends are in a better dorm and in her own dorm there is at most one friend. Thus, agent $i$ Pareto-envies $j$.

\item
Case 4. Four agents in $N^*$ are assigned to dorm 2 and the other one is assigned to dorm 1.
Suppose agent $i \in N^*$ is the one assigned to dorm 1.
Other than the four agents in $N^*$, there is another agent outside of $N^*$ who is assigned to dorm 2, so at least two agents in $\{8,9,10\}$ are assigned to dorm 1 and at least one of them, say $j$, is not a friend of $i$. 
In this case, $j$ does not share a dorm with her friend (i.e., $j-5$) and is assigned to a worse dorm.
Thus $j$ Pareto-envies $N^*\setminus\{i,j-5\}$.

\item
Case 5. Three agents in $N^*$ are assigned to dorm 1 and the other two are assigned to dorm 2.
Suppose agent $i \in N^*$ is assigned to dorm 2, which is the worse dorm for her. 
Agent $i$ has at least three friends in dorm 1 but at most two friends in dorm 2.
Thus, there is at least one agent $j$ in dorm 1 who is not a friend of $i$, and agent $i$ Pareto-envies $j$ in this assignment.

\item
Case 6. Three agents in $N^*$ are assigned to dorm 2 and the other two are assigned to dorm 1.
Assume that $\{i,j,k\} \subset N^*$ are assigned to dorm 2 and $\{l,h\} \subset N^*$ are assigned to dorm 1.
Since each dorm has capacity 5, at least one agent in $\{i+5,j+5,k+5\}$ is not in dorm 2. 
Without loss of generality, assume this is agent $i+5$. 
Then for agent $i$, she shares a worse dorm with two of her friends $j$ and $k$ while three of her friends $\{i+5,l,h\}$ are in the better dorm. 
Thus, agent $1$ Pareto-envies the other two agents in dorm 1.
\end{itemize}

Therefore, no assignment is PEF in this instance.
\end{Proof}

\begin{proposition}
\label{thm:PEF-NPhard}
If resources are allowed to have different capacities, then deciding whether a given dorm sharing instance admits a PEF assignment is NP-complete, even when there are only two resources with binary external values.
\end{proposition}

\begin{Proof}
The problem is clearly in NP: a PEF assignment is a witness of a yes-instance and can be verified in polynomial time.
To show the hardness, our approach is similar to that in the proof of Proposition~\ref{thm:EF-NP-c}; we show a reduction from the {\sc Clique} problem.

Given an instance $\langle G = (V, E), k \rangle$ of {\sc Clique}, without loss of generality, $|V|/2 < k \le |V|$ (see the proof of Proposition~\ref{thm:EF-NP-c}), we construct the following dorm sharing instance.
Let the set of agents be $N = N_1 \cup N_V$, where:
\begin{itemize}
\item 
Every agent has value $1$ for resource $1$, and value $0$ for resource $2$. Given this valuation, in any PEF assignment, each agent assigned to resource $2$ should not be able to obtain a higher external utility by swapping their position with an agent on resource $1$.

\item
$N_1$ contains a set of $2k$ agents who have external value $1$ for every other agent in $N_1$ and external value $0$ for each agent in $N_V$ unless the value is defined to be $1$ below.

\item
$N_V = \{a_i : i \in V\}$ contains $|V|$ agents, each corresponding to a vertex on $G$.
Each agent $a_i \in N_V$ has external value $1$ for each $a_j \in N_V$ such that $\{i,j\} \in E$, as well as $2k - d_i - 2$ arbitrary agents in $N_1$, where $d_i$ denotes the degree of $i$ on $G$. 
Note that since $k>|V|/2$, we have $2k - d_i - 2 \ge 0$.
All other external values are $0$.
\end{itemize}
Let the resource capacities be $c_1 = k + |V|$ and $c_2 = k$.

Observe that by the above valuations, every agent $a \in N_1$ must be assigned to resource $1$ in a PEF assignment: if they are assigned to resource $2$, their external utility is at most $k-1$ while they can get external utility at least $k$ from their remaining friends on resource $1$.
Thus, only agents in $N_V$ can be assigned to resource $2$.

It can be verified that if the {\sc Clique} instance is a yes-instance, assigning the $k$ agents corresponding to a size-$k$ clique on $G$ to resource $2$ gives a PEF assignment.
In particular, for every agent $i$ on resource $2$, their external value is $k-1$. Suppose that they swap their position with a non-friend on resource $1$. 
Their only friends on resource 1 are their $2k-d_i-2$ friends in $N_1$ and the remaining $d_i - (k-1)$ friends in $N_v$,
so their external value after the swapping is at most $2k-d_i-2 + d_i - (k-1) = k - 1$. (Note that $d_i \ge k-1$ since $i$ is in a size $k$ clique.)

Conversely, if the {\sc Clique} instance is a no-instance, no matter which $k$ agents we assign to resource $2$, some of them gets external utility at most $k-2$. Each of these agents has $2k - d_i - 2$ friends in $N_1$ and at least $\max\{0,\ d_i - (k - 2)\}$ friends in $N_V$ who are assigned to resource $1$, so swapping their position with a non-friend on resource $1$ would give them external utility at least $2k - d_i - 2 + d_i - (k - 2) = k > k -2$, which implies that no assignment can be PEF.
\end{Proof}

\section{PEF Dorm Assignment for $c=2$} 
\label{sec:dorm}

In this section, we demonstrate the existence of a PEF assignment in any dorm sharing instance 
where every dorm has capacity exactly 2 
by presenting an efficient constructive algorithm.
We first extend the notation and introduce several useful results.
Let $G=(V,E)$ be an arbitrary externality graph, where nodes represent agents and edges represent friendships.
A matching $\cM$ in $G$ is a set of edges without common nodes. It is called: 
\begin{itemize}
\item
a {\em maximum matching}, if no other matching contains more edges;
\item
a {\em perfect matching}, if the edges in it cover all the $|V|$ nodes of $G$; and 
\item
a {\em nearly perfect matching}, if the edges in it cover $|V|-1$ nodes of $G$. 
\end{itemize}

For any set of nodes $A\subseteq V$, denote by $G\setminus A$ the induced subgraph of $G$ after the nodes in $A$ and the associated edges are removed from $G$.
The subgraph $G\setminus A$ may be composed of one or more disjoint {\em components} (i.e., connected subgraphs). 
A component is called \emph{even} (or \emph{odd}) if it contains an even
(or odd) number of nodes. 
We will use the concept of {\em Tutte set} defined as follows.

\begin{definition}[\bf Tutte set]
\label{def:tutte-set}
Given a graph $G=(V,E)$,
a set $A\subseteq V$ of nodes is called a \emph{Tutte set} if every maximum matching $\cM$ of $G$ can be decomposed as
$$\cM = \cM_{\cD} \cup \cM_{\cB} \cup \cM_{A,\cD},$$
where 
the set $\cM_{\cD}$ contains a {\em nearly} perfect matching in each (odd) component of $G\setminus A$; 
the set $\cM_{\cB}$ contains a perfect matching in each (even) component of $G\setminus A$; 
and $\cM_{A,\cD}$ is a matching that matches every node in $A$ to a node in some odd component of $G\setminus A$.
Note that $G\setminus A$ might contain some unmatched vertices.
\end{definition}

\begin{figure}[t]
\begin{center}
\vspace{3mm}
\setlength{\unitlength}{0.4cm}
\begin{tikzpicture}[scale = 0.8]

\begin{scope}[thick]

\tikzstyle{my dotted} = [line width=0.8pt, line cap=round, dash pattern=on 0pt off 2pt]

\draw (0, 0.5) -- (1.5, 0.5);
\draw[my dotted] (0, 0.5) -- (0, 2);
\draw (0, 2) -- (1.5, 2);
\draw[my dotted] (1.5, 0.5) -- (1.5, 2);
\draw[my dotted] (1.5, 0.5) -- (0, 2);

\filldraw[fill=black, draw=black] (0, 0.5) circle (0.08cm);
\filldraw[fill=black, draw=black] (1.5, 0.5) circle (0.08cm);
\filldraw[fill=black, draw=black] (0, 2) circle (0.08cm);
\filldraw[fill=black, draw=black] (1.5, 2) circle (0.08cm);

\draw (2.5, 1) -- (4, 1);
\filldraw[fill=black, draw=black] (2.5, 1) circle (0.08cm);
\filldraw[fill=black, draw=black] (4, 1) circle (0.08cm);

\filldraw[fill=black, draw=black] (1, 4) circle (0.08cm);
\draw[my dotted] (0, 0.5) -- (1, 4);
\draw (2.4, 4) -- (1, 4);

\filldraw[fill=black, draw=black] (2.4, 4) circle (0.08cm);
\filldraw[fill=black, draw=black] (4, 5) circle (0.08cm);
\filldraw[fill=black, draw=black] (4, 3) circle (0.08cm);

\draw[my dotted] (2.4, 4) -- (4, 5);
\draw[my dotted] (2.4, 4) -- (4, 3);
\draw (4, 3) -- (4, 5);

\filldraw[fill=black, draw=black] (4.5, 4) circle (0.08cm);
\end{scope}

\draw[rounded corners] (-0.3, 0.2) rectangle (4.3, 2.3) {};
\draw (5.2, 1) node {$\cG_{A^-}^\eve$};

\draw[rounded corners] (0.7, 3.7) rectangle (1.3, 4.3) {};
\draw (0.2, 4) node {$A$};

\draw[rounded corners] (2.1, 2.7) rectangle (4.8, 5.3) {};
\draw (5.6, 4) node {$\cG_{A^-}^\odd$};
\end{tikzpicture}

\vspace{3mm}
\caption{An illustration of Gallai-Edmonds decomposition. The solid lines form a maximum matching and dotted lines are edges not included in this matching. 
Here $\cG_{A^-}^\odd$ contains two odd components, which contain 1 node and 3 nodes, respectively, 
and $\cG_{A^-}^\eve$ contains two even components, which contain 2 and 4 nodes, respectively.}
\label{figure:tutte}
\end{center}
\end{figure}
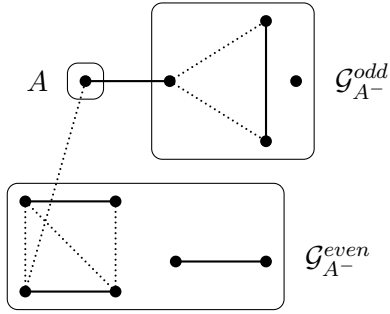

The decomposition in the above definition is called {\em Gallai-Edmonds decomposition} (see Figure \ref{figure:tutte}). Note that all the components in this decomposition are disjoint from each other in the induced graph $G\setminus A$. A nice property of Tutte sets, according to the following result, is that one such set can be computed in polynomial time.

\begin{lemma}[Gallai-Edmonds structure theorem \cite{lovasz2009matching,cheriyan1997randomized}]\label{GE-decomp}
Given a graph $G=(V,E)$, a Tutte set $A$ on $G$ can be constructed in $O(n^3)$ time.
\end{lemma}

Moreover, we will also use Hall's theorem presented below. 

\begin{lemma}[Hall's theorem \cite{hall1934representation}]
\label{lem:hall}
Given any bipartite graph $G = (L,R; E)$ with node sets $L$ and $R$, and edge set $E$ such that $|L| \leq |R|$, 
there exists a matching with size $|L|$ 
if and only if for every subset $S\subseteq L$, it holds that $|S|\leq |\cN(S)|$, where 
\[
\cN(S) = \{i \in R \mid \{i, j\} \in E \mbox{ for some } j\in S\}
\]
is the neighborhood of $S$, i.e., the set of all nodes in $R$ adjacent to some element of $S$.
\end{lemma}

When the the dorm capacity is $2$, PEF essentially requires that at least one of the following two situations holds for every agent $i$:
(1) agent $i$ shares an arbitrary dorm with one of her friends; or
(2) agent $i$ (weakly) prefers her dorm to the dorms of her friends (otherwise, $i$ would envy the agent who shares a dorm with her friend).
Hence, one possible approach to finding a PEF assignment is to compute a maximum matching in the externality graph. 
If all the agents are covered by this matching, then we pair them up according to this matching and assign each pair to an arbitrary dorm. This way the first situation will hold for all the agents.
However, if this maximum matching does not cover all agents, we need to make sure that every unpaired agent gets a better dorm than all their friends do. 
To this end, we make use of the Gallai-Edmonds decomposition. The idea is to allocate dorms that are ``bad'' for the unpaired agents to the paired ones.
We present Algorithm \ref{alg:captwo}, which computes a PEF assignment for any given instance with dorm capacity $2$. 
The correctness of this algorithm is shown via Lemmas~\ref{lmm:alg-terminates} and \ref{lmm:alg-PEF}, and Figure \ref{fig:four cases} presents four examples to illustrate the four cases in the {\bf while} loop of Algorithm~\ref{alg:captwo}.

\begin{figure}[t]
\subfloat[][Case 1]{
\begin{tikzpicture}[scale = 0.8]
\begin{scope}[thick]

\tikzstyle{my dotted} = [line width=0.8pt, line cap=round, dash pattern=on 0pt off 2pt]

\filldraw[fill=black, draw=black] (0, 0) circle (0.08cm);
\filldraw[fill=black, draw=black] (0, 1.5) circle (0.08cm);
\draw (0,0) -- (0,1.5);
\draw (-0.5, 0) node {$L$};
\draw (-0.5, 1.5) node {$A$};

\filldraw[fill=black, draw=black] (1, 0) circle (0.08cm);
\filldraw[fill=black, draw=black] (1, 1.5) circle (0.08cm);
\draw (1,0) -- (1,1.5);

\filldraw[fill=black, draw=black] (2, 0) circle (0.08cm);
\draw (2,-0.5) node {$i$};

\filldraw[fill=black, draw=black] (3, 0) circle (0.08cm);
\draw (3,-0.5) node {$i'$};

\end{scope}

\end{tikzpicture}
\label{fig:case1}
}
\hspace{16pt}
\subfloat[][Case 2]{
\begin{tikzpicture}[scale = 0.8]
\begin{scope}[thick]

\tikzstyle{my dotted} = [line width=0.8pt, line cap=round, dash pattern=on 0pt off 2pt]

\filldraw[fill=black, draw=black] (0, 0) circle (0.08cm);
\filldraw[fill=black, draw=black] (0, 1.5) circle (0.08cm);
\draw (0,0) -- (0,1.5);
\draw (-0.5, 0) node {$L$};
\draw (-0.5, 1.5) node {$A$};

\filldraw[fill=black, draw=black] (1, 0) circle (0.08cm);
\filldraw[fill=black, draw=black] (1, 1.5) circle (0.08cm);
\draw (1,0) -- (1,1.5);
\draw (0, -0.68) node {\,};

\end{scope}
\end{tikzpicture}
\label{fig:case2}
}
\hspace{16pt}
\subfloat[][Case 3]{
\begin{tikzpicture}[scale = 0.8]
\begin{scope}[thick]

\tikzstyle{my dotted} = [line width=0.8pt, line cap=round, dash pattern=on 0pt off 2pt]

\filldraw[fill=black, draw=black] (0, 0) circle (0.08cm);
\filldraw[fill=black, draw=black] (0, 1.5) circle (0.08cm);
\draw (0,0) -- (0,1.5);
\draw (-0.7, 0) node {$L$};
\draw (-0.7, 1.5) node {$A$};
\draw (0.3, 0) node {$i$};

\filldraw[fill=black, draw=black] (1, 0) circle (0.08cm);
\filldraw[fill=black, draw=black] (1, 1.5) circle (0.08cm);

\filldraw[fill=black, draw=black] (2, 0) circle (0.08cm);
\filldraw[fill=black, draw=black] (3, 0) circle (0.08cm);

\draw [red] (0,1.5) -- (1,0);
\draw [red] (1,1.5) -- (2,0);
\draw[rounded corners] (-0.3, -0.3) rectangle (2.3, 0.3) {};
\draw (2.4, 0.5) node {$\cN(S)$};

\draw[rounded corners] (-0.3, 1.2) rectangle (1.3, 1.8) {};
\draw (1.6, 1.5) node {$S$};
\draw (0, -0.68) node {\,};

\end{scope}

\end{tikzpicture}
\label{fig:case3}
}
\hspace{16pt}
\subfloat[][Case 4]{
\begin{tikzpicture}[scale = 0.8]
\begin{scope}[thick]

\tikzstyle{my dotted} = [line width=0.8pt, line cap=round, dash pattern=on 0pt off 2pt]

\filldraw[fill=black, draw=black] (0, 0) circle (0.08cm);
\filldraw[fill=black, draw=black] (0, 1.5) circle (0.08cm);
\draw (0,0) -- (0,1.5);
\draw (-0.5, 0) node {$L$};
\draw (-0.5, 1.5) node {$A$};

\filldraw[fill=black, draw=black] (1, 0) circle (0.08cm);
\filldraw[fill=black, draw=black] (1, 1.5) circle (0.08cm);
\draw (1,0) -- (1,1.5);

\draw (0,1.5) -- (1,0);
\draw (0,1.5) -- (2,0);
\draw (1,1.5) -- (2,0);
\draw (1,1.5) -- (3,0);

\filldraw[fill=black, draw=black] (2, 0) circle (0.08cm);
\draw (2,-0.5) node {$i$};
\filldraw[fill=black, draw=black] (3, 0) circle (0.08cm);
\draw (3,-0.5) node {$i'$};
\draw (3.4, 0.5) node {$\cN(S)$};

\end{scope}

\end{tikzpicture}
\label{fig:case4}
}
\caption{\label{fig:four cases} 
Illustration of the four cases in the {\bf while} loop of Algorithm~\ref{alg:captwo}. 
In Case (3), the red edges represent the nearly perfect matching $\cM'$ between $S$ and $\cN(S)$.
}
\end{figure}

\begin{algorithm}[p]
\caption{Find a PEF assignment when dorms have capacity $2$. \label{alg:captwo}}
 
\SetKwInOut{Input}{Input}
\SetKwInOut{Output}{Output}

\Input{
A dorm assignment instance $\cI=(N,M,\vv,\ee,\cc)$ with $c_i=2$ for all $i\in N$.
}

\Output{
An assignment of agents in $N$ to dorms in $M$.
}

\smallskip

\setlength{\leftmargini}{0mm}

\begin{enumerate}[start=0]
\item 
Initialization: 

Let $G = (N, E)$ be the externality graph of $\cI$;

Let $\tilde{M}$ denote the set of unassigned dorms throughout (initially, $\tilde{M} = M$).

\item
Compute a Tutte set $A \subseteq N$ of $G$, and a maximum matching $\cM$.

\item
\For{each {\bf\em even} component $X$ in $G \setminus A$}
{
	
	Since $A$ is a Tutte set, $\cM$ contains a perfect matching for $X$.
	Assign each matched pair in $X$ to an arbitrary dorm in $\tilde{M}$.
}

\item
Let $L = \emptyset$.

\For{each {\bf\em odd} component $X$ in $G \setminus A$}
{
	Since $A$ is a Tutte set, $\cM$ contains a {\em nearly} perfect matching for $X$. 
	Let $i$ be the unmatched agent in $X$ and add it into $L$.
	Assign each matched pair in $X$ to one of the $\frac{|X| - 1}{2}$ least preferred dorms of agent $i$ in $\tilde{M}$.
}

\item
Let $G^*=(A, L; E^*)$ be the bipartite graph between $A$ and $L$, such that $(a,l) \in E^*$ if and only if $a \in A$, $l\in L$, and $(a,l)\in E$.
For every $S \subseteq A$, let $\cN(S)$ denote the neighborhood of $S$ on $G^*$.

\While{$A \cup L \neq \emptyset$}
{

\tcp{See Figure \ref{fig:four cases} for an illustration of the cases below}

\uIf{{\bf(Case 1)} there exists a pair of agents $i,i'\in L$ such that $\{i,i'\} \cap e = \emptyset$ for all $e\in E^*$}
{

Assign $i$ and $i'$ to an arbitrary dorm in $\tilde{M}$. 
}
	
	 \uElseIf{{\bf(Case 2)} $|S| = |\cN(S)|$ for some (nonempty) $S \subseteq A$}
	 {
	Find a perfect matching $\cM'$ between $S$ and $\cN(S)$. 
	Assign each matched pair in $\cM'$ to an arbitrary dorm in $\tilde{M}$. 
	}

	 \uElseIf{{\bf(Case 3)} $|S| = |\cN(S)| - 1$ for some $S \subseteq A$}
	 {

	Find a nearly perfect matching $\cM'$ between $S$ and $\cN(S)$, in which all nodes in $S$ are covered. 
	Let $i \in \cN(S)$ be the unmatched agent in $\cM'$, and assign each matched pair in $\cM'$ to one of the $|S|$  {least} preferred dorms of agent $i$ in $\tilde{M}$. 
	}
	\ElseIf{{\bf(Case 4)} $|S| \le |\cN(S)| - 2$ for all (nonempty) $S \subseteq A$}
	{
	Find a pair of agents  $i,i' \in L$ who have the same most preferred dorm in $\tilde{M}$ and assign them to this most preferred dorm. 
	}
 
Remove all the assigned agents from $A$ and $L$, and remove their adjacent edges from $G^*$.
}
\end{enumerate}

\end{algorithm}

\begin{lemma}
\label{lmm:alg-terminates}
Algorithm~\ref{alg:captwo} always terminates with an assignment of $N$ to $M$.
\end{lemma}
\begin{Proof}
It suffices to show that the while loop at Step 4 always terminates.
We first argue that the following inequalities hold throughout Step 4: 
\begin{equation}
\label{eq:while-loop-key} 
|S| \le |\cN(S)|, \quad \text{ for all } S \subseteq A, 
\end{equation}
where $\cN(S)$ denotes the neighborhood of $S$ on the bipartite graph $G^*$.

Indeed, since $A$ is a Tutte set and $\cM$ is a maximum matching, according to Definition~\ref{def:tutte-set}, $\cM$ matches each node in $A$ to a node in $L$.
Thus, \eqref{eq:while-loop-key} holds before the while loop is executed.
We will argue that:
\begin{itemize}
\item[(i)]
If \eqref{eq:while-loop-key}  holds at the beginning of some round, at least one of the conditions defining Cases 1--4 must be true; 
\item[(ii)]
Moreover, no matter which case is selected, the algorithm will proceed as described, and \eqref{eq:while-loop-key}  will hold at the end of that round as long as it holds at the beginning of it.
\end{itemize}
By induction, this will imply that the while loop will continue as long as $A\cup L \neq \emptyset$. 
Since at least one pair of agents is assigned to a dorm in each of the four cases,
it follows that the while loop will indeed end with $A\cup L = \emptyset$, whereby every agent is assigned to some dorm.

\smallskip

Indeed, given that \eqref{eq:while-loop-key} holds, at least one of the conditions defining Cases 2-4 must be true, so (i) is obvious and it only remains to show (ii).
We consider the case selected by the algorithm.
\begin{itemize}
\item If it is {\bf Case 1}, since no agent in $A$ is a neighbor of $i$ or $i'$, the neighborhood $\cN(S)$ of each subset $S \subseteq A$ will not change after $i$ and $i'$ are removed from $G'$, and \eqref{eq:while-loop-key} will still hold.

\item If it is {\bf Case 2}, first, we need to argue that we can find a perfect matching as described in the algorithm. Indeed, given \eqref{eq:while-loop-key}, by Hall's theorem (Lemma~\ref{lem:hall}), there exists a matching of size $|S|$, which is a perfect matching between $S$ and $\cN(S)$ given that $|S| = |\cN(S)|$ in this case.

To see that \eqref{eq:while-loop-key} will still hold at the end of this round, suppose for the sake of contradiction that it is violated after agents in $S$ and $\cN(S)$ (and the adjacent edges) are removed from $G^*$.
In other words, we have $|Q| > |\cN(Q)\setminus \cN(S)|$ for some $Q \subseteq A \setminus S$ (where $\cN(Q)$ denotes the neighborhood of $Q$ before the removal of $\cN(S)$).
It follows that 
\begin{align*}
|Q\cup S| 
= |Q| + |S| 
&> \left|\cN(Q) \setminus \cN(S)\right| + \left|\cN(S)\right| \\
&= \left|\cN(Q) \cup \cN(S)\right| \\
&= \left|\cN(Q \cup S)\right|.
\end{align*}
Since $Q\cup S$ is a subset of $A$, this means that \eqref{eq:while-loop-key} does not hold even before $S$ and $\cN(S)$ are removed from $G^*$, which contradicts our assumption. 

\item If it is {\bf Case 3}, then similarly to Case 2, by Hall's theorem and the assumption that \eqref{eq:while-loop-key} holds at the beginning of this round, there exists a matching of size $|S|$ between $S$ and $\cN(S)$. 
Suppose for the sake of contradiction that \eqref{eq:while-loop-key} breaks after the removal of the assigned agents (in this case, these are agents in $S$ and $\cN(S) \setminus \{i\}$).
We have 
$$|Q| > \left|\cN(Q) \setminus \left(\cN(S) \setminus \{i\} \right) \right|$$
for some $Q \subseteq A \setminus S$.
Since we also have $|S| = |\cN(S)| -1$ in Case 3, it follows that
\begin{align*}
|Q \cup S| 
= |Q| + |S| 
&> \left|\cN(Q) \setminus \left(\cN(S) \setminus \{i\} \right) \right| + \left|\cN(S)\right| - 1 \\
&= \left|\cN(Q) \setminus \left(\cN(S) \setminus \{i\} \right) \right| + \left|\cN(S)  \setminus \{i\} \right|  \\
&= \left|\cN(Q) \cup \left(\cN(S) \setminus \{i\} \right) \right| \\
&\ge \left|\cN(Q \cup S)\right| - 1.
\end{align*}
Since sizes of sets are integers, this means that $|Q \cup S| \ge \left|\cN(Q \cup S)\right|$.
However, the fact that the algorithm selected Case 3, instead of Case 2, means that the condition defining Case 2 does not hold; namely, we have $|X| < |\cN(X)|$ for all subsets $X \subseteq A$. 
This contradicts the above inequality since $Q \cup S$ is a subset of $A$.

\item If it is {\bf Case 4}, we need to argue first that we can indeed find a pair of agents who have the same most preferred dorm. 
By the condition defining Case 4, we have 
$|A| \leq |\cN(A)| - 2 \le |L|-2$, which means that 
$$|L| \ge \frac{|A| + |L|}{2} +1 = |\tilde{M}| + 1.$$
Thus, there is one more agent in $L$ than the number of unassigned dorms and by the pigeonhole principle there exist two agents who prefer the same unassigned dorm the most. 
The removal of these two agents reduces $|\cN(S)|$ by at most $2$ for all $S \subseteq A$, so given that in this case we have $|S| \leq |\cN(S)| - 2$ for all $S \subseteq A$ at the beginning of the round, \eqref{eq:while-loop-key} will still hold after the removal of $i$ and $i'$.
\end{itemize}

The proof is completed by combining the above four cases.
\end{Proof}

\begin{lemma}
\label{lmm:alg-PEF}
The assignment Algorithm~\ref{alg:captwo} generates is PEF.
\end{lemma}
\begin{Proof}
Clearly, the assignment is PEF for all the agents assigned as a matched pair, each of whom shares their dorm with a friend in the assignment.
Observe that those who are not assigned as a pair only appear in $L$, so it suffices to show that the assignment is PEF for all the agents in $L$, who are assigned only in Step 4 of the algorithm. Moreover, since each agent $i \in L$ comes from a unique component in $G \setminus A$, $i$ would only Pareto-envy agents who share dorms with agents in $A$ or agents in the same component with $i$.
The way the agents in $i$'s odd component are assigned in Step 3 ensures that these agents do not get a better dorm than $i$ does, so $i$ will not Pareto-envy them.
Thus, in what follows we only need to argue that either $i$ shares a dorm with her friend in $A$ or $i$ gets a dorm that is better than any dorm assigned to her friends in $A$.


Suppose for the sake of contradiction that $i$ does not share a dorm with any of her friends and one of her friends  $i^* \in A$ gets a better dorm than $i$.

If $i$ does not share a dorm with her friends, then she can only get a dorm in some round where the
algorithm proceeds with Case 1 or 4.
Further, $i^*$ can only get a dorm in Case 2 or 3. 
If $i$ is assigned at an earlier iteration than $i^*$, then: $i^*$ is not $i$'s friend if $i$ gets assigned in Case 1, and $i$ has a better dorm than $i^*$ if $i$ gets assigned in Case 4. Both contradict the assumption. 
If $i$ is assigned after $i^*$, then $i^*$ is not $i$'s friend if $i^*$ gets assigned in Case 2, which means $i^*$ gets a dorm in Case 3 and $i$ is the only unmatched agent.
However, if this is the case, the least preferred dorms of $i$ would have been assigned to the matched agents,
which also contradicts the assumption.



Since the choice of $i$ in $L$ is arbitrary, the assignment is PEF and this completes the proof.
\end{Proof}

In fact, Algorithm~\ref{alg:captwo} can be implemented in polynomial time. 
The key to the implementation is to find an efficient way to determine whether the conditions defining Cases 2 and 3 are true or not, and to compute a subset $S$ satisfying these conditions.
We demonstrate how this can be done and summarize our results as the following key theorem.

\begin{theorem}
\label{thm:PEF-cap2}
Given any dorm assignment instance with capacity $2$ for all dorms, a PEF assignment always exists and can be computed in polynomial time.
\end{theorem}

\begin{Proof} 
Given Lemmas~\ref{lmm:alg-terminates} and \ref{lmm:alg-PEF}, we show that there is a way to implement Algorithm~\ref{alg:captwo} in polynomial time to complete this proof.
In what follows, we will use the same notation as in Algorithm~\ref{alg:captwo}. 
Thus, $G^*$ is the bipartite graph constructed in Step 4 of Algorithm~\ref{alg:captwo} and for each $S\subseteq A$, we denote the neighborhood of $S$ on $G^*$ by $\cN(S)$.

Consider each step of Algorithm~\ref{alg:captwo}.
In Step 1, a Tutte set and a maximum matching can be computed in polynomial time by Lemma \ref{GE-decomp}.
Following that, Steps 2 and 3 trivially run in polynomial time.
Thus, it suffices to argue that Step 4 can be implemented in polynomial time.

Indeed, in the while loop at Step 4, if we have determined which case to proceed with, the subsequent procedure for each case can be implemented efficiently.
Specifically, for Cases 1, the assignment procedure is trivial.
For Cases 2 and 3, given $S$, to find a perfect or nearly perfect matching it suffices to compute a maximum matching, which can be done in polynomial time as we already know.
For Case 4, to find the pair $\{i,i'\}$, we can enumerate all the $O(n^2)$ agent pairs, and the subsequent assignment procedure is trivial, too.
Therefore, we only need to show that we can efficiently determine whether the conditions defining  Cases 1--3 are true or not. 
When none of them hold, the condition defining Case 4 must be true as we have argued in the proof of Lemma~\ref{lmm:alg-terminates}.

To check if the condition defining Case 1 is true, we can simply enumerate all the agent pairs in $L$, which takes time $O(n^2)$.
To check if the condition defining Case 2 is true, we enumerate every $\ell \in L$ and apply the following procedure, which attempts to generate a (nonempty) set $S \subseteq A$ such that $|S| = |\cN(S)|$.
\begin{enumerate}
\item
We first compute a maximum matching $\cM$ on $G^*$. 
Since \eqref{eq:while-loop-key} holds as we have shown in the proof of Lemma~\ref{lmm:alg-terminates}, by Hall's theorem $\cM$ covers every agent in $A$.
For every agent $i \in A \cup L$, we let 
\[
\cM(i) = \left\{ j \in A \cup L: j \text{ is matched to } i \text{ in } \cM \right\}.
\]
Note that $\cM(i)$ is either a singleton or an empty set, and we also write $\cM(X) = \bigcup_{i\in X} \cM(i)$ for a set $X$.

\item Let $S = \cM(\ell)$.

\item 
For each $i \in \cN(S)$, we add the agent in $\cM(i)$ into $S$ if it is not in $S$ yet, and repeat this step until $|S| = |\cN(S)|$ or no new agent can be added into $S$ in this way (i.e., when $\cM(i) \subseteq S$ for all $i \in \cN(S)$).
\end{enumerate}

Clearly, the above procedure finishes in polynomial time. 
We prove its correctness next. 

\begin{claim}
\label{clm:S}
Suppose that there exists a nonempty set $S^* \subseteq A$ such that $|S^*| = |\cN(S^*)|$. Then the above procedure will successfully generate such a set for some $\ell \in L$.
\end{claim}

\begin{Proof}[Proof of Claim~\ref{clm:S}]
Since $\cM$ covers every agent in $A$, we have $|\cM(S^*)| = |S^*|$.
Now that $|S^*| = |\cN(S^*)|$, we further get that $|\cM(S^*)| = |\cN(S^*)|$; hence, $\cM(S^*) = \cN(S^*) \neq \emptyset$.

Consider an arbitrary $\ell \in \cM(S^*)$ and let $S$ be the set produced with this $\ell$ in the above procedure.
Suppose for the sake of contradiction that $|S| \neq |\cN(S)|$.
Observe that the following properties hold throughout the repetition of Step 3: 
\begin{itemize}
\item $|S| \le |\cN(S)|$; and
\item $S \subseteq S^*$ and $\cN(S) \subseteq \cN(S^*)$.
\end{itemize}
Hence, the assumption that $|S| \neq |\cN(S)|$ implies $|S| < |\cN(S)|$.
According to Step~3, this means some $i \in \cN(S)$ is not matched to any element in $S$. In other words, some $i \in \cN(S^*)$ is not matched to any element in $S^*$, which contradicts the fact that $\cM(S^*) = \cN(S^*)$ we argued above. The assumption that $|S| \neq |\cN(S)|$ cannot be true.
\end{Proof}

Similarly, to check the condition defining Case 3, 
we enumerate every agent pairs $\{\ell, \ell'\} \subseteq L$ (such that $\ell \neq \ell'$) and apply the same procedure but modify Step~3 to the following step.
\begin{enumerate}


\item[3.] 
For all $i \in \cN(S) \setminus \{\ell'\}$, we add $\cM(i)$ into $S$ if it is not in $S$ yet, and repeat this step until $|S| = |\cN(S)|-1$ or no new agent can be added into $S$ in this way.
\end{enumerate}
If there exists a set $S^*$ with $|S^*| = |\cN(S^*)|-1$ (and assume that Case 2 is not true), then there exists a pair $\{\ell, \ell'\}$ such that  $\{\ell, \ell' \} \subseteq \cN(S^*)$, $\cM(\ell) \in S^*$, and $\cM(\ell') \notin S^*$. 
Thus, by similar arguments, the above procedure will correctly generate a set $S$ with $|S| = |\cN(S)|-1$.

In summary, Algorithm~\ref{alg:captwo} generates a PEF assignment for any given instance with capacity $2$ for all dorms, and it can be implemented in polynomial time. This completes the proof.
\end{Proof}

\paragraph{Impossibility with $c \ge 3$.}

Unfortunately, if the capacity of the dorms increases to $3$, a PEF assignments may not exist.
We demonstrate this via Example~\ref{exp:no-PEF} and Proposition~\ref{prp:example-no-PEF-cap-3} below.

\begin{example}
\label{exp:no-PEF}
There are $3$ dorms and $9$ agents. 
The agents' external values are defined by the graph in Figure~\ref{figure:c>=3}; 
namely, for each $i \in \{1,3,5,7\}$, agents $i$ and $i+1$ are friends with each other.
Every agent has value $j$ for each dorm $j\in \{1,2,3\}$.
\end{example}

\begin{figure}[t]
\begin{center}
\vspace{3mm}
\setlength{\unitlength}{0.4cm}
\begin{tikzpicture}[scale = 0.85]
\begin{scope}[thick]
\draw (0, 0.5) -- (1.5, 0.5);
\draw (-0.3, 0.2) node {$1$};
\draw (1.8, 0.2) node {$2$};
\filldraw[fill=black, draw=black] (0, 0.5) circle (0.08cm);
\filldraw[fill=black, draw=black] (1.5, 0.5) circle (0.08cm);

\draw (3, 0.5) -- (4.5, 0.5);
\draw (2.7, 0.2) node {$3$};
\draw (4.8, 0.2) node {$4$};
\filldraw[fill=black, draw=black] (3, 0.5) circle (0.08cm);
\filldraw[fill=black, draw=black] (4.5, 0.5) circle (0.08cm);

\draw (6, 0.5) -- (7.5, 0.5);
\draw (5.7, 0.2) node {$5$};
\draw (7.8, 0.2) node {$6$};
\filldraw[fill=black, draw=black] (6, 0.5) circle (0.08cm);
\filldraw[fill=black, draw=black] (7.5, 0.5) circle (0.08cm);

\draw (9, 0.5) -- (10.5, 0.5);
\draw (8.7, 0.2) node {$7$};
\draw (10.8, 0.2) node {$8$};
\filldraw[fill=black, draw=black] (9, 0.5) circle (0.08cm);
\filldraw[fill=black, draw=black] (10.5, 0.5) circle (0.08cm);

\filldraw[fill=black, draw=black] (12, 0.5) circle (0.08cm);
\draw (12.3, 0.2) node {$9$};
\end{scope}
\end{tikzpicture}
\vspace{3mm}
\caption{The externality graph of Example~\ref{exp:no-PEF}.}
\label{figure:c>=3}
\end{center}
\end{figure}

\begin{proposition}
\label{prp:example-no-PEF-cap-3}
The instance in Example~\ref{exp:no-PEF} does not admit any PEF assignment.
\end{proposition}

\begin{Proof}
Suppose for the sake of contradiction that there is a PEF assignment.
Then for any agent $i \neq 9$ who is assigned to dorm 3, the friend of agent $i$ has to be assigned to dorm 3 as well; otherwise, this friend will Pareto-envy the other agents in dorm 3. 
Since the capacity of dorm 3 is $3$, after assigning agent $i$ and her friend in this dorm, we can only assign agent $9$ to fill up the dorm as every other agent has a friend. 
Thus, the other three pairs of agents are assigned to dorms 1 and 2, which means that at least one pair of friends must be assigned to two different dorms.
As a result, for this pair of friends, the one in dorm 1 will Pareto-envy the other agents in dorm 2 because they share a better dorm with her friend. 
Therefore, no PEF assignment exists for this instance.
\end{Proof}

\section{Utility Guarantees of EF and PEF Assignments}
\label{sc:EF-PROP}

We have investigated the existence and computation of EF and PEF assignments. 
In this section, we shift our focus to the utility guarantees of these assignments.

\subsection{Utility Guarantee of EF Assignments}

We start with EF assignments
and focus on the dorm sharing model.
Recall that in the dorm sharing model, all the resources have the same capacity $c$, and we have $n=c\cdot m$.
Note that $c$ is not necessarily $2$ in this section.
The results we will present compare utilities (sum of internal and external values) of EF assignments with a utility threshold $\Delta_i$ for each agent $i$.
The threshold, defined in \eqref{eq:PROP}, resembles the agent's utility guarantee in a {\em proportional assignment}, which is another extensively studied concept in the literature of fair division.
\begin{align}
\label{eq:PROP}
\Delta_i = \frac{1}{m} \sum_{j\in M} v_{ij} + \frac{c-1}{n-1} \cdot \sum_{\ell \in N\setminus\{i\}}e_{i \ell}.
\end{align}
Specifically, in classical non-sharable resource allocation settings (where agents are assigned disjoint bundles of resources), 
a proportional assignment is one in which every agent gets at least $\frac{1}{n}$ of the total value of the resources according to their own valuation. Imagine that an agent can legally demand to sell all the resources and divide the proceeds equally among the agents.
Hence, for internal values, the first part of $\Delta_i$, as defined in \eqref{eq:PROP}, is $\frac{1}{n}$ of the total value $\sum_{j\in M}c \cdot v_{ij}$ if all the agents have the same valuation as agent $i$.
For external values, since every agent $i$ shares their resource with $c-1$ mates and their average value for a friend is $\frac{1}{n-1}\sum_{\ell \in N \setminus\{i\}} e_{i\ell}$, which is the second part of \eqref{eq:PROP}.
Note that the denominator is $n-1$ instead of $n$ since $i$ has at most $n-1$ friends.
In other words, $\Delta_i$ is the expected utility of agent $i$ if resources are assigned uniformly at random.
A weaker version of proportionality, where the denominator is replaced by $n$, is discussed at the end of this subsection.



It is straightforward that an assignment that ensures $\Delta_i$ for every agent $i$ may not be EF.
This is true even in the special setting where the agents have no external values.
As a simple example, when there are 3 resources and $n=3c$ agents, if agent 1 has values $v_{11} = 1$, $v_{12} = 0$, and $v_{13} = 2$ for the resources while all the other agents have value $1$ for every resource, assigning agent $1$ to resource $1$ and the others agents arbitrarily but feasibly ensures $\Delta_i$ for all agents but it is not EF for agent $1$.
Nevertheless, as we show in Proposition~\ref{thm:EFToPROP:identical}, every EF assignment {\em almost} guarantees $\Delta_i$ for every agent $i$, leaving only a small gap of $\frac{\Delta_i}{n}$.


\begin{proposition}
\label{thm:EFToPROP:identical}
For any dorm assignment instance with $c \ge 2$ and any EF assignment $X$ in this instance, it holds that $u_i(X_i) \ge (1-\frac{1}{n}) \cdot \Delta_i$ for all agents $i$.
\end{proposition}

\begin{Proof}
Suppose that $X$ is an EF assignment.
Consider an arbitrary agent $i$ and let $r\in M$ be the resource assigned to $i$.
Then, 
\begin{align}\label{eq:ef-prop:j_i}
u_i(X) =  v_{ir} + \sum_{j \in X_{r}} e_{ij} \ge  v_{ir} + \frac{c-1}{c} \cdot \sum_{j \in X_{r}} e_{ij},
\end{align}
Since $X$ is EF, for any resource $j \in M \setminus \{r\}$ and any agent $i'\in X_j$, we have
\begin{align}
u_i(X)  \ge u_i \left(X^{i \exch i'}\right) =  v_{ij} + \sum_{\ell \in X_j \setminus \{i'\}} e_{i \ell}. \label{eq:ef-prop:j}  
\end{align}
Summing up \eqref{eq:ef-prop:j} for all $i'\in X_j$ and dividing both sides of the inequality by $c$ gives
\begin{align}\label{eq:ef-prop:Xj}
u_i(X) & \ge v_{ij} + \frac{c-1}{c} \cdot\sum_{\ell \in X_j} e_{i\ell} 
\end{align}
Now summing up \eqref{eq:ef-prop:j_i} and \eqref{eq:ef-prop:Xj} for all resources $j \neq r$ and dividing both sides of the inequality by $m$ gives

\begin{align}
u_i(X) &\ge \frac{1}{m} \cdot \sum_{j\in M} v_{ij} + \frac{1}{m} \cdot  \frac{c - 1}{c} \cdot \sum_{j}\sum_{\ell \in X_j} e_{i\ell} \nonumber \\
& = \frac{1}{n} \cdot \sum_{j\in M} c \cdot v_{ij} + (c - 1) \cdot \frac{1}{n} \cdot \sum_{\ell \in N }e_{i\ell} & (\text{recall $n = c\cdot m$}) \label{eq:prop:ef:n-1} \\
&\ge \frac{n-1}{n} \left( \frac{1}{n} \cdot \sum_{j\in M} c \cdot v_{ij} + \frac{c - 1}{n-1} \cdot \sum_{\ell \in N}e_{i\ell} \right) 
 = \frac{n-1}{n} \cdot  \Delta_i, \nonumber 
\end{align}
which completes the proof of the lemma.
\end{Proof}


The above lower bound of $(1 - \frac{1}{n}) \cdot \Delta_i$ is essentially tight according to the following result.

\begin{proposition}
For any constant $\delta > 0$, there exists an instance of the dorm sharing model and an EF assignment $X$, such that $u_i(X) < \left( 1 -  \frac { 1} {n} + \delta \right) \cdot \Delta_i$ for some agent $i$.
\end{proposition}

\begin{Proof}
Suppose that there are $m$ resources, which need to be assigned to $n = c\cdot m$ agents,
where $c \ge 2$ is the capacity of every resource. 
Let $X=(X_1, \cdots, X_m)$ be an assignment where agent 1 is assigned to resource 1, i.e., $1 \in X_1$.
Let the value of agent 1 for resource 1 be $v_{11} = c - 1$ and for every other resource $j \ge 1$ be $v_{1j} = 0$.
Let the external value of agent 1 be $e_{1\ell} = 0$ for every $\ell \in X_1$, and $e_{1\ell} = 1$ for every other agent $\ell \notin X_1$.

Thus, 
$u_1(X) = v_{11} = c-1$,
and exchanging agent 1 with any agent in a resource $j>1$,
will not increase the utility of agent 1; $X$ is envy-free.
Let us compute $\Delta_1$.
By construction of the instance, we have 
\begin{align*}
\Delta_1 &=
\frac{1}{m} \sum_{j\in M} v_{ij} + \left(\frac{n}{m} - 1 \right) \cdot \left( \frac{1}{n-1} \sum_{\ell \in N }e_{i\ell} \right) \\
&= \frac{c-1}{m} + (c-1) \cdot \frac{c(m-1)}{cm-1} 
= u_1(X) \cdot \left( \frac{1}{m} + \frac{c(m-1)}{cm-1} \right) \\
&= u_1(X) \cdot \frac{n^2 - c}{ n^2 - n}.
\end{align*}
Hence, when $n/c$ is sufficiently large, for any constant $\delta > 0$, we have
\[
u_1(X) =  \frac { n^2 - n} {n^2 - c} \cdot \Delta_1 
= \left( 1 -  \frac { n - c} {n^2 - c} \right) \cdot \Delta_1
< \left( 1 -  \frac { 1} {n} + \delta \right) \cdot \Delta_1.
\]
The assignment $X$ cannot ensure utility better than $\left(1 - \frac{1}{n} + \delta \right)\cdot \Delta_i$ for every agent $i$.
\end{Proof}

Finally, we remark that if the denominator in the right-hand term of Equation \eqref{eq:PROP} is replaced by $n$, i.e., 
\begin{align*}
\Delta_i' = \frac{1}{m} \sum_{j\in M} v_{ij} + \frac{c-1}{n} \cdot \sum_{\ell \in N}e_{i \ell},
\end{align*}
then all EF assignments ensure that every agent $i$ has utility at least $\Delta_i'$. 
This can be seen from the fact that Equation \eqref{eq:prop:ef:n-1} in the proof of Proposition \ref{thm:EFToPROP:identical} is exactly $\Delta_i'$.

\subsection{Utility Guarantee of PEF Assignments}

We next move to PEF assignments. 
If we modify the average measures of the internal and external values in Equation \eqref{eq:PROP}, we are able to obtain Proposition~\ref{lem:PEF-PPROP} for two special settings of the dorm sharing model. 
Essentially, PEF implies a property that resembles the Pareto frontier of the notion of $\Delta_i$, where for every agent at least one of her internal and external values satisfies the corresponding part in $\Delta_i$.
In the following Proposition~\ref{lem:PEF-PPROP}, the first condition means that the resource agent $i$ gets is ranked among her top 50\% of all resources.
The second condition is a direct modification of the external value in Equation \eqref{eq:PROP} that rounds the value to its nearest integer.
We denote by $\lfloor x \rceil$ the nearest integer of a number $x$, i.e., $\lfloor x \rceil = \lceil x \rceil$ if $x \ge \lfloor x \rfloor + \frac{1}{2}$, and $\lfloor x \rceil = \lfloor x \rfloor$, otherwise.

\begin{proposition}
\label{lem:PEF-PPROP}
For any dorm sharing instance with $c = 2$ or $m = 2$, if an assignment $X$ is PEF, then for every agent $i$ at least one of the following conditions holds:
\begin{enumerate}
\item 
$\left|\{j\in M \setminus \{r_i(X)\}: v_{i r_i(X)} \ge v_{ij}\} \right| \ge \frac{1}{2} m$; or
\item $e_i(X) \ge \left\lfloor \frac{c-1}{n-1} \sum_{\ell \in N}e_{i\ell} \right\rceil$.
\end{enumerate}
\end{proposition}

\begin{Proof}
We first consider the case where $c = 2$ (and $m$ is arbitrary).
In this case $m = n/2$.
Consider an arbitrary agent $i$ and the following two possibilities with respect to the value $\sum_{\ell \in N }e_{i \ell}$. 

\begin{itemize}
\item
 If $\sum_{\ell \in N }e_{i \ell} < m$, then we have
\[
\left\lfloor \frac{c-1}{n-1} \sum_{\ell \in N }e_{i\ell} \right \rceil 
\le \left \lfloor \frac{1}{n-1} \cdot \left(\frac{n}{2} - 1\right)\right \rceil = 0.
\]
Thus the second condition holds for agent $i$.

\item 
If $m \leq \sum_{\ell \in N }e_{i\ell} \le 2m-1$,
then we have 
\[
\left\lfloor \frac{c-1}{n-1} \sum_{\ell \in N }e_{i\ell} \right\rceil = 1.
\]
Hence, as long as agent $i$ shares a resource with a friend of hers, the second condition will hold.
On the other hand, if agent $i$ does not have a friend in $X$, then given that $\sum_{\ell \in N }e_{i\ell} \ge m$ and now the agents have binary external values, the friends of agent $i$ would occupy at least $\lceil\frac{m}{2}\rceil$ resources in $X$.
Since assignment $X$ is PEF and agent $i$ now envies every agent who shares a dorm with one of her friends, the resource agent $i$ gets must have a higher value for agent $i$ than all those occupied by her friends; hence, the first condition holds for agent $i$.
\end{itemize}

Next we consider the case where $m = 2$ (and $c$ is arbitrary).
In this case $X = (X_1, X_2)$ and each resource is shared among $\frac{n}{2}$ agents.
Consider an arbitrary agent $i$ and without loss of generality we can assume that $i\in X_1$.
Suppose that condition 1 does not hold for agent $i$; since now there are only two resources, this means that $v_{i1} < v_{i2}$.
We show that condition 2 must hold in this case if $X$ is PEF for agent $i$. 
Indeed, if $X$ is PEF for agent $i$ while $v_{i1} < v_{i2}$, it must be that agent $i$ does not envy any other agent for their external values, i.e, $e_i(X) \ge e_i(X^{i \exch \ell})$ for all $\ell \in N$.
Consider the following situations.

\begin{itemize}
\item
If all the $\frac{n}{2}$ agents in $X_2$ are friends of agent $i$, then all the other $\frac{n}{2} - 1$ agents in $X_1$ must also be friends of agent $i$ as otherwise, agent $i$ would envy every agent in $X_2$ for their external value.
This implies that $\sum_{\ell \in N }e_{i\ell} = n-1$
and $e_i(X) = \frac{n}{2} - 1$;
consequently,
$$\left\lfloor \frac{c -1}{n-1} \sum_{\ell \in N}e_{i\ell} \right\rceil = \left\lfloor \frac{\frac{n}{2} -1}{n-1} \sum_{\ell \in N}e_{i\ell} \right\rceil = \frac{n}{2} - 1 = e_i(X),$$
so the second condition holds. 

\item
If some agent $j \in X_2$ is not a friend of agent $i$,
then agent $i$ must have as many friend in $X_1$ as in $X_2$ (otherwise, $e_i(X) < e_i(X^{i \exch j})$).
It follows that
$$e_i(X) \ge \left\lceil \frac{1}{2} \sum_{\ell \in N }e_{i\ell} \right\rceil
\ge \left\lfloor \frac{c -1}{n-1} \sum_{\ell \in N}e_{i\ell} \right\rceil,$$
so the second condition holds, too.
\end{itemize}

Since the choice of agent $i$ is arbitrary, this completes the proof.
\end{Proof}

We cannot hope to derive the same result beyond the above two settings.
The following example shows that Proposition \ref{lem:PEF-PPROP} does not hold when $c = 3$ and $m= 3$.

\begin{example}
\label{exp:no-PPROP}
Let $c=3$ and $m=3$ (so there are $n=c\cdot m = 9$ agents).
Agents 1, 2, 3, and 4 are friends with each other,
agents 5 and 6 are friends with each other, and agents 7, 8, and 9 are friends with each other (see the externality graph in Figure \ref{figure:example:pef-pprop}).
Suppose all the agents have the same value over the resources: with value $1$ for resource 1, value $2$ for resource 2, and value $3$ for resource 3.
\end{example}

The assignment $X=(X_1, X_2, X_3)$ with $X_1=\{1,2,3\}$, $X_2=\{4,5,6\}$ and $X_3 = \{7,8,9\}$ is PEF:
other than agent 4, every agent either gets their favorite resource, or share a resource with a friend;
agent 4 on the other hand gets a better resource than all of her friends.
However, $X$ does not satisfy the conditions in Proposition \ref{lem:PEF-PPROP} for agent 4. 
The resource she gets is more valuable than only one of the other resources, while $m/2 = 3/2 > 1$ in this case, so the first condition does not hold. Meanwhile, her external value is $0$, so the second condition does not hold, either.

\begin{figure}[htbp]
\begin{center}
\setlength{\unitlength}{0.6cm}
\vspace{3mm}
\begin{tikzpicture}[scale = 1]
\begin{scope}[thick]
\draw (0, 0.5) -- (1.5, 0.5);
\draw (0, 0.5) -- (1.5, 2);
\draw (0, 0.5) -- (0, 2);
\draw (0, 2) -- (1.5, 2);
\draw (1.5, 0.5) -- (1.5, 2);
\draw (1.5, 0.5) -- (0, 2);

\draw (-0.3, 0.2) node {$1$};
\draw (-0.3, 2.2) node {$2$};
\draw (1.8, 0.2) node {$4$};
\draw (1.8, 2.2) node {$3$};

\filldraw[fill=black, draw=black] (0, 0.5) circle (0.08cm);
\filldraw[fill=black, draw=black] (1.5, 0.5) circle (0.08cm);
\filldraw[fill=black, draw=black] (0, 2) circle (0.08cm);
\filldraw[fill=black, draw=black] (1.5, 2) circle (0.08cm);

\draw (3.5, 1) -- (5, 1);
\filldraw[fill=black, draw=black] (3.5, 1) circle (0.08cm);
\filldraw[fill=black, draw=black] (5, 1) circle (0.08cm);

\draw (3.2, 0.8) node {$5$};
\draw (5.3, 0.8) node {$6$};

\draw (7, 0.5) -- (9, 0.5);
\draw (7, 0.5) -- (8, 2);
\draw (9, 0.5) -- (8, 2);

\draw (6.7, 0.2) node {$7$};
\draw (9.2, 0.2) node {$8$};
\draw (8.3, 2.2) node {$9$};

\filldraw[fill=black, draw=black] (7, 0.5) circle (0.08cm);
\filldraw[fill=black, draw=black] (9, 0.5) circle (0.08cm);
\filldraw[fill=black, draw=black] (8, 2) circle (0.08cm);
\end{scope}
\end{tikzpicture}
\vspace{3mm}
\caption{The externality graph of Example~\ref{exp:no-PPROP}.}
\label{figure:example:pef-pprop}
\end{center}
\end{figure}

\section{Conclusion and Discussion}
\label{sc:conclusion}

In this paper, we study a resource sharing problem with externalities, in which we consider both the agents' values for the resources and their external values for other agents.
We studied EF assignments in this model, the existence of such assignments, and the computation.
In general, an EF assignment may not exist.
Only in a special setting where all resources have the same capacity $2$ and all agents have dichotomous preferences over other agents, an EF assignment is guaranteed to exist under the notion of Pareto EF and can be computed efficiently.
The existence guarantee is invalidated even when we slightly generalize this setting.
We also investigate the utility guarantee of EF and PEF assignments.

There are several interesting future directions of this work.
In our paper, we only focus on the case of unit-demand agents so a natural extension is the case when agents demand multiple resources.
It would also be interesting to adopt other ways to soften the strong requirements of EF, such as EF up to one or more items \cite{budish2011combinatorial}.
One may also consider other popular fairness notions in the literature
such as the maximin share fairness \cite{budish2011combinatorial} and extend these notions to the setting with externalities.

Other variants of the notions we studied in this work may also be of interest in future work.
For example, we can relax the EF notion and define a weaker variant in which every agent only compares their situation with the worst agent on every resource, rather than with every agent as required by the EF notion.
This leads to the wEF notion defined below.

\begin{definition}[\bf wEF assignment]
\label{def:wef}
An assignment $X$ is {\em weak envy-free (wEF)} if 
for every agent $i \in N$ and every resource $r \in M$, there exists $j \in X_{r}$ such that $u_i(X) \ge u_i(X^{i \exch j})$.
\end{definition}

On the other hand, our PEF notion is a rather weak notion as it only requires one of internal and external utilities does not lead to envy.
Imagine if all the dorms are identical for one agent, then every assignment is PEF for this agent as if she does not care about her dormmates at all.
Therefore, a stronger requirement, standing between EF/wEF and PEF, might be necessary, and one possibility is the following strong PEF (sPEF) notion.

\begin{definition}[\bf sPEF assignment]
\label{def:sPEF}
An assignment $X$ is {\em strong Pareto envy-free (sPEF)} if for every pair of agents $i,j \in N$ at least one of the following three conditions holds:
\begin{enumerate}
\item $v_i(X) > v_i(X^{i\exch j})$; 
\item $e_i(X) > e_i(X^{i \exch j})$; or
\item $v_i(X) = v_i(X^{i\exch j})$ and $e_i(X) = e_i(X^{i \exch j})$.
\end{enumerate}
\end{definition}

In other words, agent $i$ envies $j$ if and only if the vector $(v_i(X^{i\exch j}),e_i(X^{i\exch j}))$ Pareto-dominates $(v_i(X),e_i(X))$.
Unfortunately, neither a wEF nor a sPEF assignment always exists, even in the special case of dorm sharing model with $c = 2$ and $m = 2$ (recall that a PEF assignment always exists in this case). This can be seen via the following example.

\begin{example}
\label{ex:def:alternative}
Let $N=\{1,2,3,4\}$ and $M=\{1,2\}$.
The two dorms are identical to all the agents and both have a capacity of $2$.
Agents $\{1,2,3\}$ are friends of each other (i.e., a 3-clique), and agent 4 is not a friend of any other agent (i.e., a singleton). 
Then in any assignment, one agent $i \in \{1,2,3\}$ must share a dorm with agent $4$ and will thus envy every agent in the other dorm. By definition, the assignment is not wEF, nor is it sPEF as the dorms are identical but the agent's external value is strictly smaller than the other two agents in $\{1,2,3\}$.
\end{example}


Despite the non-existence results, we believe that the related computational and algorithmic problems of the above notions are interesting directions for future work.

\vskip 0.2in
\bibliography{dorm}
\bibliographystyle{apalike}

\end{document}